\documentclass[11pt,letterpaper]{article}
\usepackage[margin=1in]{geometry}
\usepackage[T1]{fontenc}
\usepackage{lmodern,amsmath,amssymb,amsthm,mathtools,microtype}
\usepackage{alltt}
\usepackage[hidelinks]{hyperref}
\hypersetup{pdftitle={Optimal Matrix Completion},pdfauthor={Emaan Hariri}}
\newcommand{\repo}{https://github.com/ehariri/optimal-matrix-completion}
\newcommand{\rev}{v1.0.0}
\newcommand{\auditbase}{\repo/blob/\rev/audit}
\newcommand{\R}{\mathbb R}
\newcommand{\E}{\mathbb E}
\newcommand{\Prob}{\mathbb P}
\newcommand{\norm}[1]{\lVert #1\rVert}
\newcommand{\op}{\mathrm{op}}
\newtheorem{theorem}{Theorem}[section]
\newtheorem{lemma}[theorem]{Lemma}

\theoremstyle{definition}

\numberwithin{equation}{section}
\title{Optimal Matrix Completion}
\author{Emaan Hariri}
\date{November 2026}
\begin{document}
\maketitle
\begin{abstract}
Convex relaxation achieves the information-theoretic sampling rate for exact matrix completion. Under standard incoherence and independent Bernoulli sampling, we prove that nuclear-norm minimization attains the minimax rate of \(\Theta(\mu rn\log n)\) expected observations for \(n\times n\) matrices of rank \(r\) and incoherence at most \(\mu\) when \(\mu r\log n=o(n)\).

Our upper bound also applies to rectangular matrices. Nuclear-norm minimization exactly recovers any fixed real \(n_1\times n_2\) matrix of rank \(r\) and incoherence at most \(\mu\) with high probability from \(O(\mu r\max\{n_1,n_2\}\log(n_1+n_2))\) expected observations. In the square case, this removes the extra \(\log(\mu r)\) factor in Ding and Chen's guarantee. The proof uses trace moments on a single observation mask to control both the tangent-space sampling error and a finite polynomial certificate. For the lower bound, we adapt Cand\`es and Tao's missed-row argument to give a matching bound for every randomized estimator.

Both bounds are verified in Lean, with the proofs and audit scripts available on GitHub \mbox{(\url{https://github.com/ehariri/optimal-matrix-completion/tree/v1.0.0})}. The formalization and this paper were developed with assistance from GPT 6.2 Astra, Claude Fable 5.2, and Gemini 4 (Carbon and Neon).
\end{abstract}

\section{Recovery from one observation mask}
Fix a real \(n_1\times n_2\) matrix \(M\) of rank \(r\ge1\). In its compact singular value decomposition \(M=U\Sigma V^{\mathsf T}\), the two frames have \(r\) orthonormal columns. We say \(M\) is \(\mu\)-incoherent if
\[
  \max_i\|U_{i,:}\|_2^2\le\mu r/n_1,\qquad
  \max_j\|V_{j,:}\|_2^2\le\mu r/n_2.
\]
Include each entry independently with probability \(p\) in the observed set \(\Omega\). Given its observed values, consider the convex program
\begin{equation}\label{eq:program}
  \min_{X\in\R^{n_1\times n_2}}\|X\|_*
  \quad\text{subject to}\quad X_{ij}=M_{ij}\quad((i,j)\in\Omega).
\end{equation}
Here \(\|X\|_*\) is the sum of the singular values, and \(\|X\|_{\op}\) is the largest singular value.

\begin{theorem}\label{thm:main}
There is an absolute constant \(C>0\) such that the following holds. Let \(n_1,n_2,r\) be positive integers, \(\mu\ge1\), and \(D=n_1+n_2\). For each fixed \(\mu\)-incoherent \(n_1\times n_2\) matrix \(M\) of rank \(r\) and each \(p\in[0,1]\) satisfying
\[
 p\ \ge\ C\,\frac{\mu r\log D}{\min(n_1,n_2)},
\]
the probability that \(M\) is the unique minimizer of \eqref{eq:program} is at least \(1-D^{-3}\).
\end{theorem}

We take \(C=2^{38}\) throughout. All unqualified logarithms are natural. At the threshold, the expected sample count is \(C\mu r\max(n_1,n_2)\log D\).

\paragraph{Earlier work.}
Cand\`es and Recht proved exact recovery by nuclear-norm minimization \eqref{eq:program} using a dual certificate \cite{cr}. Cand\`es and Tao expanded the inverse in that certificate as a Neumann series and bounded its terms by a moment method. Their theorems assume a strong incoherence property and carry further logarithmic factors \cite[Thms.~1.1 and~1.2]{ct}. Gross introduced the golfing scheme, which builds a certificate from independent batches, and Recht applied it to matrix completion \cite{gross,recht}. With it Chen obtained recovery of an \(n\times n\) matrix from order \(\mu rn\log^2n\) observations under the two incoherence bounds above alone \cite{chen}.

In their conference paper, Balcan et al.~\cite[Thm.~5]{blwz} stated an \(O(\mu rD\log D)\) bound for unique completion with high probability among matrices of rank at most \(r\) with the same incoherence bounds, under uniform sampling without replacement. The authors later removed this claim after identifying a problem with its proof \href{https://arxiv.org/pdf/1704.08683v5\#page=10}{(arXiv version~5, footnote~4)}.

Ding and Chen returned to one mask with a leave-one-out argument, started by \(O(\log(\mu r))\) rounds of sample splitting. Their bound is of order \(\mu r\log(\mu r)\,n\log n\), and they note that the factor \(\log(\mu r)\) is absent from the lower bound \cite[Thm.~2, \S3.3.1 and \S4.3]{dc}. For square matrices Theorem~\ref{thm:main} removes that factor. For \(\mu r\log n=o(n)\), Appendix~\ref{app:minimax} proves the matching minimax lower bound by assigning independent signs to the rows of a sparse input.

We use a finite Neumann sum as the certificate. Its tangent error decays geometrically, and trace moments control the size of its perpendicular component.

The obstacle to using one mask throughout is that each successive power in the certificate uses the same random operator. In a large trace power an observed entry can therefore recur both within one factor and across several factors. A partition of the draw positions records which entries must agree. If the partition has \(b\) blocks among \(s\) positions, its deterministic coefficient gains \(L^{s-b}\), where
\[
       L=\mu r/n_1+\mu r/n_2 .
\]
The random weight of a block of size \(\ell\) costs a factor \(p^{-(\ell-1)}\) times a combinatorial coefficient. Centering kills singleton blocks, so every surviving partition has \(s-b\ge s/2\). After counting the possible blocks, the small parameter is \(sL/p\). Taking \(s\) proportional to \(\log D\) keeps that parameter small under the sampling hypothesis and makes the trace moment large enough to overcome its dimension factor.

Section~\ref{sec:certificate} reduces recovery to two estimates. Section~\ref{sec:trace} applies the same cycle estimate in the rank space for the polynomial and the ambient matrix space for the sampling error. Appendix~\ref{app:frame-start} proves this estimate and counts the coordinate classes using a connected graph. Appendix~\ref{app:mom-start} sums the trace moments. Appendix~\ref{app:duality} proves the duality inequality used for uniqueness. Appendix~\ref{app:minimax} gives the matching lower bound in the square case, and Appendix~\ref{app:lean} gives the formally verified statement.

\section{Reduction to two estimates on one mask}\label{sec:certificate}

Following the dual-certificate strategy of Cand\`es and Recht \cite[Lemma~3.1]{cr},
we use injectivity of sampling on the tangent space and a certificate supported
on the observed entries. Their certificate matches \(UV^{\mathsf T}\) exactly on
the tangent space. We use Gross's approximate criterion \cite[Sec.~II-E]{gross},
which Recht used for entry sampling \cite[\S4]{recht} and Chen used
in the Bernoulli model \cite[Prop.~2 and Appendix~A]{chen}. Near-isometry bounds the
tangent part of a feasible perturbation, so the nuclear-norm gain on its
perpendicular part can absorb a small tangent residual.

Write \(P_U=UU^{\mathsf T}\), \(P_V=VV^{\mathsf T}\), and \(S=UV^{\mathsf T}\).
Use the Frobenius inner product \(\langle X,Z\rangle=\operatorname{tr}(X^{\mathsf T}Z)\) on the matrix space.
For its subspace
\[
  T=\{UC^{\mathsf T}+BV^{\mathsf T}:C\in\R^{n_2\times r},\
                B\in\R^{n_1\times r}\}
\]
the orthogonal projections are
\[
 P_{T^\perp}Z=(I-P_U)Z(I-P_V),\qquad
 P_TZ=P_U Z+ZP_V-P_U ZP_V .
\]
Indeed, \(T^\perp=\{Z:U^{\mathsf T}Z=0,\ ZV=0\}\) by testing against \(C,B\). The first displayed map is self-adjoint, is idempotent, and has this range. In particular both projections contract the Frobenius norm. The formula for \(P_{T^\perp}\) also gives
\(\norm{P_{T^\perp}Z}_{\op}\le\norm{Z}_{\op}\).

Let \(P_\Omega\) denote entrywise restriction with zeros off \(\Omega\).
The sampling hypothesis gives \(p>0\), since \(D=n_1+n_2\ge2\) and
\(\mu,r\ge1\). At \(p=1\) feasibility determines every entry almost surely.
For every \(0<p\le1\) write \(R=p^{-1}P_\Omega\),
\(A=P_T R|_T:T\to T\), and \(\Delta=I_T-A\).
We use the Frobenius-induced norm on maps \(T\to T\).
For the matrix unit \(E_e\) at entry \(e\), put \(\phi_e=P_TE_e\).
If \(Z\in T\), then \(Z_e=\langle \phi_e,Z\rangle\) and
\[
 (A-I_T)Z=\sum_e \left(p^{-1}\mathbf 1_{e\in\Omega}-1\right)
                \langle \phi_e,Z\rangle \phi_e .
\]
The same sum defines the operator \(\Xi\) of Lemma~\ref{lem:endpoint-coeff} on the full matrix space. It vanishes on \(T^\perp\), so its norm equals its norm on \(T\).

For \(K=\lfloor\log_4 D\rfloor\), define the supported certificate
\[
 Y_K=R\sum_{\ell=0}^{K-1}\Delta^\ell S .
\]
This truncates the Neumann expansion of the Cand\`es--Recht inverse
certificate \cite[(4.2)]{cr} given by Cand\`es and Tao \cite[\S3.3]{ct}.
The last stage of Ding and Chen's construction uses the same fixed-mask sum,
starting from the residual left by independent batches \cite[(83)--(85)]{dc}.
We start from \(S\) and use one mask throughout.

Since \(S\in T\) and \(A=I_T-\Delta\), summing adjacent powers gives
\(P_TY_K=A\sum_{\ell=0}^{K-1}\Delta^\ell S=S-\Delta^K S\).
Its perpendicular component is
\begin{equation}\label{eq:certpoly}
 Q_K=P_{T^\perp}Y_K=\sum_{j=1}^K F_j,\qquad
 B_j=(R-I)\Delta^{j-1}S,\qquad F_j=P_{T^\perp}B_j .
\end{equation}
Indeed \(P_{T^\perp}\Delta^{j-1}S=0\) for each \(j\).

The first estimate below is the standard tangent-space near-isometry at
\(p\) of order \(\mu r\log n/n\) in the square case
\cite[Lemma~1]{chen}, \cite[Prop.~2]{dc}. We obtain it from the same trace
argument that controls \(Q_K\).

\begin{theorem}[Bounds on the sampling error and the cutoff]\label{thm:tails}
Under the hypotheses of Theorem~\ref{thm:main}, let
$Q_K$ be the polynomial in \eqref{eq:certpoly}.
Then
\[
 \Prob\!\left(\norm{\Delta}_{T\to T}>1/16\right)
     \le \tfrac12 D^{-3},\qquad
 \Prob\!\left(\norm{Q_K}_{\op}>3/8\right)
     \le D^{-9}/15.
\]
Here $\Delta=I_T-A$, and its norm is induced by the Frobenius norm on $T$.
\end{theorem}
We prove these two estimates in Appendix~\ref{app:mom-start}. When \(p=1\), \(R=I\) almost surely and both bad events have probability zero.

\begin{lemma}[Finite polynomial certificate]\label{lem:finite-certificate}
Under the sampling hypothesis of Theorem~\ref{thm:main}, suppose that
\(\norm{\Delta}_{T\to T}\le 1/16\) and \(\norm{Q_K}_{\op}\le 3/8\).
Every feasible perturbation \(H\) satisfies
\[
 \norm{M+H}_*-\norm{M}_*\ \ge\ \tfrac12\norm{P_{T^\perp}H}_*,
\]
and the increase is strict whenever \(H\ne0\).
\end{lemma}
\begin{proof}
A feasible perturbation \(H\) preserves the observed entries, so \(P_\Omega H=0\).
Write \(H=H_T+N\), where \(H_T=P_TH\) and \(N=P_{T^\perp}H\).
Feasibility gives \(RH=0\), hence \(AH_T=-P_TRN\). The reverse
triangle inequality and the bound on \(\Delta\) yield
\[
 \tfrac{15}{16}\norm{H_T}_F\le\norm{AH_T}_F
 \le\norm{RN}_F\le p^{-1}\norm N_F .
\]
Thus \(\norm{H_T}_F\le (2/p)\norm N_F\), and \(N=0\) forces \(H=0\).

The tangent residual \(E_K=S-P_TY_K=\Delta^K S\) satisfies
\(\norm{E_K}_F\le 16^{-K}\sqrt r\), since \(S\) has \(r\) unit singular
values. To compare it with \(p\), put \(n_{\min}=\min(n_1,n_2)\). The rate gives
\(pn_{\min}\ge 2^{38}\mu r\log D\ge64r\). As \(r\ge1\) and \(p\le1\), we
have \(D\ge64\), \(K\ge3\), and \(\sqrt r/p\le r/p\le n_{\min}/64\le D\).
Then \(D<4^{K+1}\) gives
\[
 \frac{2}{p}\norm{E_K}_F\le2D\,16^{-K}\le 8\,4^{-K}\le\tfrac18.
\]

Since \(Y_K\) is supported on \(\Omega\), \(\langle Y_K,H\rangle=0\).
Splitting this pairing along \(T\) and \(T^\perp\) gives
\(\langle S,H\rangle=\langle E_K,H_T\rangle-\langle Q_K,N\rangle\).
By Cauchy--Schwarz, the first term has absolute value at most
\(\norm{E_K}_F(2/p)\norm N_F\le\tfrac18\norm N_*\), using
\(\norm N_F\le\norm N_*\). Spectral/nuclear duality bounds the second
by \(\tfrac38\norm N_*\) in absolute value. The nuclear-norm inequality
in Appendix~\ref{app:duality} now gives
\[
 \norm{M+H}_*-\norm M_*\ge\langle S,H\rangle+\norm N_*
 \ge (1-\tfrac18-\tfrac38)\norm N_* .
\]
For \(H\ne0\) we have \(N\ne0\), proving strictness.
\end{proof}

\begin{proof}[Proof of Theorem~\ref{thm:main} assuming Theorem~\ref{thm:tails}]
Lemma~\ref{lem:finite-certificate} gives unique recovery on the intersection
of the two good events. By Theorem~\ref{thm:tails}, their combined failure
probability is at most \(\tfrac12 D^{-3}+D^{-9}/15\le D^{-3}\).
\end{proof}

\section{Following a sampled entry through a trace}\label{sec:trace}

As in the moment proof of Wigner's semicircle law \cite{tao-semicircle},
we expand a trace and use centering to discard terms in which an independent
draw appears just once. Cand\`es and Tao used trace moments to estimate the
terms of a Neumann-series certificate \cite[\S3.6]{ct}. They group the trace
terms by exact row and column equality patterns \cite[\S\S4.2--4.3]{ct}.
Here the tangent projection couples entries, and one Bernoulli mask supplies
all powers. The cumulant expansion imposes equality within each block and
allows different blocks to use the same entry. That freedom lets us apply
the frame identities to the signed coefficient sums before taking absolute
values.
Here is the coefficient bound we need. For a trace with \(s\) draw positions,
let \(C_\pi\) sum the deterministic coefficients subject to equality within
each block of a partition \(\pi\). Lemma~\ref{lem:part-count} bounds the
expected trace by \(K_0\theta^s\) if
\[
 |C_\pi|\le K_0 L^{s-|\pi|}\quad\text{for every partition without singletons},
 \qquad 8sL/p\le\theta^2,\quad 0<\theta\le1.
\]
Each block of size \(\ell\) contributes \(\ell-1\) to the exponent of \(L\),
offsetting the power \(p^{-(\ell-1)}\) in its cumulant.

Let \(\mathcal E=[n_1]\times[n_2]\), let \(\xi_e=p^{-1}\mathbf 1_{e\in\Omega}-1\), and write \(\mathcal P(e,f)=\langle E_e,P_T E_f\rangle\). For \(e=(a,b)\), write \(S_e=(UV^{\mathsf T})_{ab}\) for the scalar entry of \(S\). With \(B_j\) as in \eqref{eq:certpoly}, iteration of \(\Delta=-P_T(R-I)|_T\) gives
\begin{equation}\label{eq:wordraw}
 B_j=\sum_{e_1,\ldots,e_j\in\mathcal E}
 \left(\prod_{i=1}^j\xi_{e_i}\right)E_{e_1}
 \left(\prod_{i=1}^{j-1}\big[-\mathcal P(e_i,e_{i+1})\big]\right)S_{e_j}.
\end{equation}
Indeed \(R-I\) is diagonal with entries \(\xi_e\), while the \((e,f)\) entry of \(P_T\) in the matrix-unit basis is \(\mathcal P(e,f)\). The empty product for \(j=1\) gives \((R-I)S\).

Fix \(k\ge1\). The trace of \((B_j^{\mathsf T}B_j)^k\) bounds
\(\norm{B_j}_{\op}^{2k}\). Index its factors by \(g=(i,\epsilon)\) in
\(G=\{0,\ldots,k-1\}\times\{0,1\}\), with \(\epsilon=0\) for a
transposed factor. A map \(e:I\to\mathcal E\) assigns an entry \(e_{g,q}\)
to each of the \(s=2kj\) positions in \(I=G\times\{1,\ldots,j\}\).
Write \(E_{i,\epsilon}=E_{e_{(i,\epsilon),1}}\) for the matrix unit at the
first position of that factor. Multiplying out \eqref{eq:wordraw} gives
\[
\begin{aligned}
 \operatorname{tr}((B_j^{\mathsf T}B_j)^k)
    &=\sum_{e:I\to\mathcal E}c(e)\prod_{(g,q)\in I}\xi_{e_{g,q}},\\
 c(e)&=\operatorname{tr}\!\left(\prod_{i=0}^{k-1}E_{i,0}^{\mathsf T}E_{i,1}\right)
       \prod_{g\in G}\left[S_{e_{g,j}}
          \prod_{q=1}^{j-1}\big[-\mathcal P(e_{g,q},e_{g,q+1})\big]\right].
\end{aligned}
\]
For a partition \(\pi\) of \(I\), the coefficient to be estimated is
\[
 C_\pi=\sum_{\substack{e:I\to\mathcal E\\ e\text{ constant on each block of }\pi}}c(e).
\]
Thus each block chooses one entry, and the choices of different blocks are unrestricted.

\subsection{Closing the polynomial trace}

Write \(u_a=U_{a,:}^{\mathsf T}\), \(v_b=V_{b,:}^{\mathsf T}\).
These are two Parseval frames in the same space \(\R^r\):
\(\sum_a u_a u_a^{\mathsf T}=\sum_b v_b v_b^{\mathsf T}=I_r\).
Their squared norms are at most \(L\). We will use both frames in one
trace, even when \(n_1\ne n_2\).

\begin{lemma}[Rectangular trace coefficient]\label{lem:contraction}
Under the hypotheses of Theorem~\ref{thm:main}, let \(\pi\) partition the
\(s=2kj\) positions of \(\operatorname{tr}((B_j^{\mathsf T}B_j)^k)\)
into \(b\) blocks, with \(j,k\ge1\). In the expansion \eqref{eq:wordraw},
strip off the random scalars and sum the remaining coefficient over all
assignments of entries to blocks. This sum \(C_\pi\) satisfies
\[
                    |C_\pi|\le n_2\,3^{2k(j-1)}L^{s-b}.
\]
\end{lemma}
\begin{proof}
We use the same identity twice. For \(x_i,w_i\in\R^n\) and \(y_i,z_i\in\R^m\),
\begin{equation}\label{eq:alternating-outers}
 \operatorname{tr}\prod_{i=0}^{k-1}(x_i y_i^{\mathsf T})(z_i w_i^{\mathsf T})
   =\prod_{i=0}^{k-1}\langle y_i,z_i\rangle\langle w_i,x_{i+1}\rangle,
 \qquad x_k=x_0 .
\end{equation}
Multiplication within a pair gives the scalar \(\langle y_i,z_i\rangle\) and the outer product
\(x_i w_i^{\mathsf T}\). Multiplication between pairs gives \(\langle w_i,x_{i+1}\rangle\),
with the last inner product supplied by the trace.

Call the first position of each factor its \emph{root}. If the paired roots are
\(E_{a_i b_i}\) and \(E_{c_i d_i}\), then
\begin{equation}\label{eq:even-roots}
 \operatorname{tr}\prod_{i=0}^{k-1} E_{a_i b_i}^{\mathsf T}E_{c_i d_i}
 =\prod_{i=0}^{k-1}\mathbf1_{a_i=c_i}\,\mathbf1_{d_i=b_{i+1}},
 \qquad b_k=b_0 .
\end{equation}
This is \eqref{eq:alternating-outers} with the coordinate vectors
\(x_i=e_{b_i}\), \(y_i=e_{a_i}\), \(z_i=e_{c_i}\), \(w_i=e_{d_i}\).
For example, the four units \(E_{ab}^{\mathsf T} E_{ac}
E_{dc}^{\mathsf T} E_{db}\) have trace one for every \(a,b,c,d\).
A row is identified within each pair and a column across each pair boundary,
including the last.

The inner edges in \eqref{eq:wordraw} supply a second set of identifications.
The formula for \(P_T\) gives
\begin{align}
 \mathcal P((a,b),(a',b'))={}&
 \langle u_a,u_{a'}\rangle\mathbf1_{b=b'}
 +\mathbf1_{a=a'}\langle v_b,v_{b'}\rangle \notag\\[-2pt]
 &-\langle u_a,u_{a'}\rangle\langle v_b,v_{b'}\rangle .
 \label{eq:physical-kernel}
\end{align}
Choose one of the three terms on every edge of every factor. In the
first choice only the row coordinate contributes a Gram inner product,
in the second only the column does, and in the third both do. Call the
coordinates contributing an inner product \emph{active} on that edge.
In \(-\mathcal P\) the three choices have signs \(-,-,+\), respectively,
so each choice over all edges has a fixed overall sign.

Take one row symbol and one column symbol for each block of \(\pi\).
On each axis impose all root equalities of \eqref{eq:even-roots} on that
axis and identify the ends of every edge on which that axis is inactive.
Denote the resulting sets of equivalence classes by \(Q_U,Q_V\).
An assignment satisfying all indicators is exactly one choice of a row
for each class in \(Q_U\) and a column for each class in \(Q_V\). Choices
for distinct classes may coincide. This bijection accounts for every
surviving summand without a multiplicity.

For a fixed assignment let \(w_x\) be the frame vector selected at
class \(x\in Q_U\sqcup Q_V\), and write \(P_x=w_xw_x^{\mathsf T}\).
For an ordered list \(A\) of classes put \(P(A)=\prod_{x\in A}P_x\),
with \(P(\varnothing)=I_r\). In factor \(g\), write \(\rho_g,\gamma_g\)
for the row and column classes of its root. Direct each edge from position
\(q\) to \(q+1\), away from the root. On either axis its \emph{target}
is the class at position \(q+1\). The lists \(\mathsf{row}_g,\mathsf{col}_g\)
record, in that order of traversal, the targets of the edges active on the
respective axes. The scalar for factor \(g\), after removing the overall sign, is
\begin{equation}\label{eq:mode-dot}
 \alpha_g=\bigl(w_{\rho_g}^{\mathsf T}P(\mathsf{row}_g)\bigr)
          \bigl(w_{\gamma_g}^{\mathsf T}P(\mathsf{col}_g)\bigr)^{\mathsf T}.
\end{equation}
Indeed, multiplying \(w_x^{\mathsf T}\) by \(P_y\) takes the inner
product with \(w_y\) and leaves \(w_y^{\mathsf T}\) at the next class.
An inactive edge keeps the same class. The final inner product in
\eqref{eq:mode-dot} is precisely \(S_{ab}=\langle u_a,v_b\rangle\)
at the last physical entry.

The root identities \eqref{eq:even-roots} give
\(\rho_{i,0}=\rho_{i,1}\) and \(\gamma_{i,1}=\gamma_{i+1,0}\),
with pair indices taken modulo \(k\).
For a fixed pair abbreviate its subscripts to \(0,1\). Concatenate the lists
\[
 \gamma_0,\quad \mathsf{col}_0,\quad \overleftarrow{\mathsf{row}_0},\quad
 \rho_1,\quad \mathsf{row}_1,\quad \overleftarrow{\mathsf{col}_1}
\]
in pair order to form \(W\). An arrow reverses a list, so
\(P(\overleftarrow A)=P(A)^{\mathsf T}\) since every \(P_x\) is symmetric. Then
\begin{equation}\label{eq:mode-trace}
                       \prod_g\alpha_g=\operatorname{tr}P(W).
\end{equation}
To see the equality, read the trace as the cyclic product of inner products
of consecutive vectors, as in \eqref{eq:alternating-outers}. From
\(\gamma_0\) it follows the column path of factor 0 outward and its row
path backward to \(\rho_1=\rho_0\). This gives \(\alpha_0\) in reverse
order. From \(\rho_1\) it follows the row path of factor 1 outward and its
column path backward to the next pair's column root, giving \(\alpha_1\).
Empty paths leave the corresponding root in place.

\emph{Example with two blocks.} Take \(k=1,\ j=2\). Block \(A\) consists of
both roots and block \(B\) of both ends. Thus \(s=4,\ b=2\), and the root
trace is \(\operatorname{tr}(E_e^{\mathsf T}E_e)=1\). Before choosing kernel terms,
\(C_\pi=\sum_{e,f\in\mathcal E}\mathcal P(e,f)^2 S_f^2\), including \(e=f\).
Write \(x,x'\) for the row classes of \(A,B\).

If both edges use the first term of \eqref{eq:physical-kernel}, their column
symbols merge into one class \(y\). Thus \(Q_U=\{x,x'\}\), \(Q_V=\{y\}\).
In both factors the roots are \((x,y)\), the row list is \((x')\), and
the column list is empty. Assign \(u_a,u_c,v_b\) to \(x,x',y\). Then
\[
 W=(y,x',x,x'),\qquad
 \operatorname{tr}P(W)=\langle u_a,u_c\rangle^2\langle u_c,v_b\rangle^2.
\]
Summing first over \(a,b\) and using Parseval for each frame gives
\(\sum_c\|u_c\|^4\le L\sum_c\|u_c\|^2=rL\). The four visits use three classes.

If both edges use the third term, the column classes \(y,y'\) of \(A,B\)
also stay distinct. Both factors have roots \((x,y)\) and target lists
\((x')\), \((y')\). Assign \(v_b,v_d\) to \(y,y'\). Now
\[
\begin{aligned}
 W&=(y,y',x',x,x',y'),\\
 \operatorname{tr}P(W)&=\langle v_b,v_d\rangle^2
       \langle u_a,u_c\rangle^2\langle u_c,v_d\rangle^2 .
\end{aligned}
\]
Summing over the root indices \(a,b\) gives
\(\sum_{c,d}\|u_c\|^2\|v_d\|^2\langle u_c,v_d\rangle^2
 \le L^2\sum_{c,d}\langle u_c,v_d\rangle^2=rL^2\).
Here six visits use four classes. Both computations give the power of \(L\)
predicted by the general cycle estimate below.

Let \(t=|W|\), \(h=|Q_U|+|Q_V|\), and let \(\delta\) count edges active
on both axes. The cycle visits every class, and Appendix~\ref{app:frame-start}
gives the trace bound and the two counts
\[
 \left|\sum_{\text{class assignments}}\operatorname{tr}P(W)\right|
       \le r L^{t-h},\qquad t=s+\delta,\quad h\le b+1+\delta.
\]
The rate condition gives \(0<L\le 2p/(2^{38}\log D)<1\).
Since \(r=\sum_b\|v_b\|^2\le n_2L\), we obtain
\(rL^{t-h}\le n_2 L^{t-h+1}\le n_2 L^{s-b}\).
There are \(2k(j-1)\) edges. Summing the bounds for their
\(3^{2k(j-1)}\) choices proves the lemma.
\end{proof}

In Lemma~\ref{lem:contraction}, the factor \(3^{2k(j-1)}\) counts the three kernel terms at each of \(j-1\) edges in each of the \(2k\) factors. Taking a \(2k\)-th root leaves \(3^{j-1}\) for one factor.

\subsection{The tangent-space trace}

\begin{lemma}[Tangent trace coefficient]\label{lem:endpoint-coeff}
Under the hypotheses of Theorem~\ref{thm:main}, let \(\phi_e=P_T E_e\) in the ambient \(n_1n_2\)-dimensional Hilbert space, and set \(\Xi=\sum_e \xi_e \phi_e \phi_e^{\mathsf T}=P_T(R-I)P_T\), with \(\phi_e\) vectorized. For any partition of the \(s=2k\) positions in \(\operatorname{tr}(\Xi^{2k})\) into \(b\) blocks of size at least two, the corresponding summed trace coefficient has absolute value at most \(n_1n_2 L^{s-b}\).
\end{lemma}
\begin{proof}
The matrix units form an orthonormal basis for the Frobenius inner product. Since \(P_T\) is an orthogonal projection, for every matrix \(Z\),
\[
 \sum_e |\langle \phi_e,Z\rangle|^2
    =\sum_e |\langle E_e,P_T Z\rangle|^2
    =\|P_T Z\|_F^2\le\|Z\|_F^2 .
\]
Thus the projected units satisfy Bessel's inequality with bound one. At \(e=(i,j)\), the formula for \(P_T\) gives
\[
 \|\phi_e\|_F^2=\langle E_e,P_T E_e\rangle
 =(P_U)_{ii}+(P_V)_{jj}-(P_U)_{ii}(P_V)_{jj}\le L .
\]
Read the \(s\) positions in trace order and write down the block label at each position. Assigning an entry \(e\) to a block puts the same rank-one factor \(\phi_e\phi_e^{\mathsf T}\) at every occurrence of its label. The summed coefficient is exactly the cycle in Lemma~\ref{lem:cycle-frame}, with word length \(s\), \(b\) distinct labels and ambient dimension \(n_1n_2\). Its bound is \(n_1n_2 L^{s-b}\) for arbitrary block sizes. Centering will eliminate singleton blocks when we take the expectation.
\end{proof}

The coefficient bounds leave only scalar probability and counting. Appendix~\ref{app:mom-start} shows how the powers of \(L\) absorb the cost of repeated Bernoulli draws and how to choose a trace order that works all the way to the cutoff \(K\).
\appendix
\section{A trace bound for both estimates}\label{app:frame-start}

\subsection{One leverage factor for each repeated class}

Both traces are sums of products of rank-one matrices. In the sampling trace each label chooses a projected matrix unit. In the polynomial trace each label chooses a row of \(U\) or \(V\). In either case the map taking a vector to its inner products with all the choices has norm at most one.

\begin{lemma}[Trace bound for a cycle of frame vectors]\label{lem:cycle-frame}
Let \(W=(x_1,\ldots,x_t)\) visit every label in a finite set \(Q\) of
size \(h\), and let \(L\ge0\). For each \(x\in Q\) let \(I_x\) be finite and suppose the vectors
\((w_{x,a})_{a\in I_x}\) in \(\R^d\) satisfy
\[
       \sum_{a\in I_x}|\langle w_{x,a},v\rangle|^2\le\|v\|^2\quad(v\in\R^d),
       \qquad \|w_{x,a}\|^2\le L\quad(a\in I_x).
\]
Under an assignment \(y\in\prod_x I_x\), put
\(P_x(y)=w_{x,y_x}w_{x,y_x}^{\mathsf T}\). Then
\[
 \left|\sum_y\operatorname{tr}\big(P_{x_1}(y)\cdots P_{x_t}(y)\big)\right|
 \le d L^{t-h}.
\]
\end{lemma}
\begin{proof}
Give each label \(x\) a coordinate with values in \(I_x\cup\{\ast\}\),
where \(\ast\) denotes an empty coordinate. Work throughout on the orthogonal
direct sum of copies of \(\R^d\) indexed by these coordinate tuples. We use the
matrix-unit construction of Albar, Junge and Zhao \cite[\S2]{ajz}.
For the coordinate of \(x\), let \(E_{ab}\) send basis vector \(b\) to \(a\)
and kill the other basis vectors. Write \(P_a=w_{x,a}w_{x,a}^{\mathsf T}\).
For a label with at least two visits, insert the following operators in the
written order of the word, acting as the identity on the other coordinates.
\[
 \underbrace{\sum_a E_{\ast a}\otimes P_a}_{\rm first},\qquad
 \underbrace{\sum_a E_{aa}\otimes P_a}_{\rm interior},\qquad
 \underbrace{\sum_a E_{a\ast}\otimes P_a}_{\rm last}.
\]
For a singleton insert \(E_{\ast\ast}\otimes\sum_a P_a\). Expand the product
and take its block from the all-empty tuple to itself. Matrix units attached to
one coordinate multiply to zero unless all its visits carry the same \(a\).
For each assignment that survives, their product on that coordinate is
\(E_{\ast\ast}\), and the rank-one factors appear in their original order.
The block is therefore exactly \(\sum_y P_{x_1}(y)\cdots P_{x_t}(y)\),
even when visits to different labels interleave.

Bessel's inequality gives \(\sum_aP_a\preceq I_d\), and
\(P_a^2=\|w_{x,a}\|^2P_a\preceq L P_a\). The squared norm of the first
operator is \(\|\sum_a P_a^2\|_{\op}\le L\). The last is its adjoint.
The interior operator is block diagonal, of norm at most \(L\), and a singleton
has norm at most one. A label appearing \(m\ge2\) times thus costs
\(\sqrt L\,L^{m-2}\sqrt L=L^{m-1}\). Multiplication over all labels gives
\(L^{t-h}\). Compression to the all-empty block can only decrease the norm,
and its trace is taken in \(\R^d\), giving the factor \(d\).
\end{proof}

\subsection{Counting coordinate classes by connectivity}\label{app:class-count}

Return first to the two-block example in Lemma~\ref{lem:contraction}.
With only rows active, the block edges \(xy,x'y\) connect all three
classes. Here \(\delta=0,\ t=4=s+\delta\) and \(h=3=b+1+\delta\).
With both axes active, the block edges \(xy,x'y'\) are disjoint. Each
of the two inner steps adds an edge \(xx'\). One such edge already connects
the two block edges. Now \(\delta=2,\ t=6=s+\delta\) and
\(h=4\le b+1+\delta=5\). This is why the vertex count is an upper bound.

For general edge choices, every row class
appears either as a row root or as the target of a row-active edge:
along a factor, the class stays unchanged until such an edge is crossed.
Every block occurs in some factor, so this accounts for all row classes.
The same argument applies to the column classes. The list \(W\) retains
every active target. It retains only one root of each equal row-root
pair and one of each equal column-root pair. Thus it covers all \(h\)
classes. Each of the \(k\) pairs contributes two roots to \(W\).
Each of the \(2k(j-1)\) edges contributes one target, plus a second
target if it is active on both axes. Consequently
\[
                  t=2k+2k(j-1)+\delta=s+\delta .
\]

Make each row class and each column class a vertex. Each of the \(b\)
partition blocks contributes an edge between its row and column vertices.
Consecutive roots share a vertex by \eqref{eq:even-roots}, so all root
edges lie in one connected component. Follow any factor outward from its
root. On a step active on just one axis, the two block edges meet at the
vertex of the unchanged coordinate. On a step active on both axes, join
the two row vertices by an additional edge. This connects the consecutive
block edges even when both coordinates change.

Every block occurs on one of these paths, and every class is incident
to a block edge. The resulting graph is therefore connected, with \(h\)
vertices and at most \(b+\delta\) edges. A connected graph on \(h\)
vertices needs at least \(h-1\) edges, so \(h\le b+\delta+1\).
Together with \(t=s+\delta\), this gives \(t-h+1\ge s-b\).

\section{From trace contractions to the two probability estimates}
\label{app:mom-start}\label{sec:moments}

Recall that $\xi_e=\mathbf 1_{e\in\Omega}/p-1$ for each physical entry $e$.
Different entries have independent draws. Every occurrence of the same
entry in a polynomial uses the same draw. We first account for these
repetitions exactly and then use the deterministic trace contractions.

\subsection{Summing the costs of repeated entries}

Assume $0<p\le1$ and write $m_\ell=\E\xi_e^\ell$, with $m_0=1$.
Then $m_1=0$, $m_2=(1-p)/p$, and $|\xi_e|\le 1/p$, so $|m_\ell|\le p^{-\ell}$
for every $\ell\ge0$. For $\ell\ge2$ we gain one power of $p$ by retaining the second moment:
\[
 |m_\ell|\le \E[|\xi_e|^{\ell-2}\xi_e^2]\le p^{-(\ell-2)}m_2\le p^{-(\ell-1)}.
\]
The following one-variable recursion is the scalar form of the classical
moment--cumulant partition identity \cite{speed}. It defines the cumulants by
\[
 m_{a+1}=\sum_{i=0}^{a}\binom ai\kappa_{i+1}m_{a-i}
 \qquad(a\ge0).
\]
The last summand is $\kappa_{a+1}$. In particular $\kappa_1=0$, and
\begin{equation}\label{eq:cumulant-cap}
 |\kappa_{a+1}|\le a!\,(2/p)^a.
\end{equation}
Indeed, for $a\ge1$, solve for $\kappa_{a+1}$ and apply the triangle inequality. Put $\tau=a!p^{-a}$. The moment term $|m_{a+1}|$ is at most
$\tau$. The induction hypothesis for $\kappa_{i+1}$, the bound $|m_{a-i}|\le p^{-(a-i)}$, and
$\binom ai i!=a!/(a-i)!\le a!$ give a bound $2^i\tau$ for the $i$th
summand. The total is at most $(1+\sum_{i=0}^{a-1}2^i)\tau=2^a\tau$.

Fix a finite set \(I\) of draw positions. A term indexed by \(\alpha\) has signed
weight \(w_\alpha\) and entry labels \(g_\alpha(i)\). For a partition \(\pi\)
of \(I\), let \(C_\pi(w)\) be the sum of the weights for which \(g_\alpha\) is
constant on each block. For the polynomial trace in Section~\ref{sec:trace},
the term index \(\alpha\) is the map \(e:I\to\mathcal E\), its weight is \(c(e)\),
and \(g_e(i)=e(i)\). Thus \(C_\pi(w)=C_\pi\), including assignments that give
different blocks the same entry.

\begin{lemma}[Removing one cumulant block]\label{lem:part-count}
Suppose \(8|I|L/p\le\theta^2\), \(0<\theta\le1\), \(L,K_0\ge0\), and
\(|C_\pi(w)|\le K_0 L^{|I|-|\pi|}\) for every partition whose blocks have
size at least two. Then
\[
 \left|\sum_\alpha w_\alpha\,\E\prod_{i\in I}\xi_{g_\alpha(i)}\right|
 \le K_0\theta^{|I|}.
\]
\end{lemma}
\begin{proof}
For \(I=\varnothing\), the hypothesis for the empty partition gives the result.
Otherwise distinguish \(i\in I\) and write \(n=|I|-1\).
For fixed \(\alpha\), independence factors the joint moment by entry.
Apply the scalar cumulant recursion to the positions carrying the same entry as \(i\).
In its binomial coefficient, choosing \(q\) partners means choosing a subset
\(B\subset I\setminus\{i\}\) of size \(q\) whose labels equal \(g_\alpha(i)\).
This gives the factor \(\kappa_{q+1}\) times the joint moment on the complement
of \(\Gamma=\{i\}\cup B\). The singleton term vanishes.

For each nonempty \(B\), restrict \(w_\alpha\) to those terms with constant label
on \(\Gamma\), setting the other weights to zero. Every partition of the
remaining positions into blocks of size at least two extends to one of \(I\)
by adjoining \(\Gamma\). Its coefficient for the restricted weights is exactly
the coefficient of that extended partition. Consequently the hypothesis holds
on the complement with constant \(K_0L^q\). This keeps the signs in each
coefficient sum and permits later blocks to use the same entry as \(\Gamma\).
By induction the contribution for this \(B\) has absolute value at most
\(|\kappa_{q+1}|K_0 L^q\theta^{n-q}\).

Summing over the choices of partners and using \eqref{eq:cumulant-cap} gives
\[
 K_0\sum_{q=1}^{n}\binom nq q!(2L/p)^q\theta^{n-q}
 \le K_0\theta^{n+1}\sum_{q=1}^{n}2^{-q}
 \le K_0\theta^{n+1}.
\]
Indeed \(\binom nq q!\le n^q\), \(2nL/p\le\theta^2/2\), and
\(\theta^{n+q}\le\theta^{n+1}\) for \(q\ge1\).
\end{proof}

\subsection{Choosing the trace power}

Let \(\mathcal B\) be a homogeneous matrix polynomial of degree \(j\) in the
draws, with \(\nu\) columns. The expansion of
\(\operatorname{tr}((\mathcal B^{\mathsf T}\mathcal B)^k)\) has \(s=2kj\)
draw positions per term. Use the block sums \(C_\pi\) of Section~\ref{sec:trace}
with the coefficients of this trace. Suppose that for some \(\lambda\ge0\)
every partition without singletons satisfies
\[
 |C_\pi|\le \nu\lambda^{2k} L^{s-|\pi|}.
\]
Lemma~\ref{lem:part-count} with \(K_0=\nu\lambda^{2k}\) bounds the expected
trace by \(\nu\lambda^{2k}\theta^s\) when \(8sL/p\le\theta^2\) and
\(0<\theta\le1\). On each mask the trace is the sum of the \(2k\)-th
powers of the singular values, so it bounds \(\norm{\mathcal B}_{\op}^{2k}\).

Under the hypotheses of Theorem~\ref{thm:main}, the sampling lower bound
and $p\le1$ imply $p>0$ and
\begin{equation}\label{eq:width-budget}
 \frac{L}{p}\log D\le 2^{-37},\qquad
 L=\frac{\mu r}{n_1}+\frac{\mu r}{n_2}.
\end{equation}
Indeed, $L\le 2\mu r/\min(n_1,n_2)$. Also $D\ge2$ and
$\min(n_1,n_2)\ge2^{38}\mu r\log D\ge2^{38}\log 2$,
so $\log D\ge1$ and~\eqref{eq:width-budget} implies $L\le1$.
For \(1\le j\le\log D\) any integer \(k\) with
\(4\log D<kj\le8\log D\) suffices. The formal proof takes the smallest
power of two with \(kj>4\log D\). Its predecessor \(k/2\) satisfies
\((k/2)j\le4\log D\), giving the upper bound. Thus \(8\log D<s\le16\log D\), and
\[
 8s L/p\le128\,(L/p)\log D\le 2^{-30}\le(1/256)^2 .
\]
Take $\theta=1/256$ in Lemma~\ref{lem:part-count}.
As $8\log 4\ge10$, we have $4^{-s}\le D^{-10}$. Splitting
$256^{-s}=4^{-s}64^{-s}$ therefore gives
\begin{equation}\label{eq:dyadic-moment}
 \E\norm{\mathcal B}_{\op}^{2k}
 \le \nu D^{-10}\bigl(\lambda\,64^{-j}\bigr)^{2k}.
\end{equation}

\begin{proof}[Proof of Theorem~\ref{thm:tails}]
The ambient operator $\Xi=P_T(R-I)P_T$ of Lemma~\ref{lem:endpoint-coeff}
restricts to $A-I_T=-\Delta$ on $T$ and vanishes on $T^\perp$.
Apply~\eqref{eq:dyadic-moment} to $\Xi$ with
\(j=1\), \(\lambda=1\), and \(\nu=n_1n_2\) as in Lemma~\ref{lem:endpoint-coeff}.
Markov's inequality at $1/16$ gives a failure probability at most
$n_1n_2 D^{-10}(16/64)^{2k}\le \tfrac12 D^{-3}$, using
$n_1n_2\le D^2/4$ and $D\ge2$.

For \(1\le j\le K\le\log D\), apply Lemma~\ref{lem:contraction} and
\eqref{eq:dyadic-moment} to \(B_j=(R-I)\Delta^{j-1}S\) with
\(\lambda=3^{j-1}\) and \(\nu=n_2\le D\). Since \(3^{j-1}\le4^j\) and
\(P_{T^\perp}\) contracts the operator norm, the half-order \(k=k_j\ge1\)
selected for this degree satisfies
\[
 \E\norm{F_j}_{\op}^{2k}\le\E\norm{B_j}_{\op}^{2k}
    \le D^{-9}(16^{-j})^{2k}.
\]
Markov's inequality at \(4^{-j}\) then gives
\[
 \Prob\!\left(\norm{F_j}_{\op}>4^{-j}\right)
 \le D^{-9}\bigl(16^{-j}/4^{-j}\bigr)^{2k}
 \le D^{-9}16^{-j}.
\]
If every degree satisfies its bound, the triangle inequality gives
\(\norm{Q_K}_{\op}\le\sum_{j=1}^K4^{-j}\le1/3<3/8\).
A union bound over these degrees therefore bounds the probability of
\(\norm{Q_K}_{\op}>3/8\) by \(D^{-9}\sum_{j=1}^K16^{-j}\le D^{-9}/15\).
\end{proof}
\section{A nuclear-norm inequality for feasible perturbations}\label{app:duality}

For a real rectangular matrix \(Z\) with compact singular value decomposition
\(Z=\sum_i s_i a_i b_i^{\mathsf T}\), every \(W\) satisfies
\[
 |\langle W,Z\rangle|
   \le\sum_i s_i\,|a_i^{\mathsf T}Wb_i|
   \le \norm W_{\op}\sum_i s_i .
\]
The matrix \(W_0=\sum_i a_i b_i^{\mathsf T}\) has operator norm at most one
and \(\langle W_0,Z\rangle=\sum_i s_i=\norm Z_*\). For \(Z=0\) take
\(W_0=0\). This proves the spectral/nuclear dual formula, including
attainment for rectangular matrices.

Apply this construction to \(Z=P_{T^\perp}H\). For \(s_i>0\) the identities
\(a_i=s_i^{-1}Zb_i\) and \(b_i=s_i^{-1}Z^{\mathsf T}a_i\) put the singular
vectors in the ranges of \(Z\) and \(Z^{\mathsf T}\), respectively. Since
\(Z=(I-P_U)H(I-P_V)\), these ranges lie in \(\operatorname{col}(U)^\perp\)
and \(\operatorname{col}(V)^\perp\). Thus the corresponding \(W_0\) belongs to
\(T^\perp\). The maps \(S=UV^{\mathsf T}\) and \(W_0\) have orthogonal
initial spaces and orthogonal ranges. For any \(x\in\R^{n_2}\),
\[
 \begin{aligned}
 \norm{(S+W_0)x}_2^2
 &=\norm{S P_Vx}_2^2+\norm{W_0(I-P_V)x}_2^2
 \\
 &\le\norm{P_Vx}_2^2+\norm{(I-P_V)x}_2^2=\norm x_2^2.
 \end{aligned}
\]
Use \(S+W_0\) as a dual test matrix for \(M+H\). Since \(M\in T\) and
\(W_0\in T^\perp\), we have \(\langle W_0,M\rangle=0\). The compact SVD gives
\(\langle S,M\rangle=\operatorname{tr}\Sigma=\norm M_*\), and the choice of
\(W_0\) gives \(\langle W_0,H\rangle=\langle W_0,P_{T^\perp}H\rangle
=\norm{P_{T^\perp}H}_*\). Expanding the pairing yields
\[
 \norm{M+H}_*\ge \norm M_*+\langle S,H\rangle+\norm{P_{T^\perp}H}_* .
\]

\clearpage
\section{The square minimax lower bound}\label{app:minimax}
The missed-row argument of Cand\`es and Tao \cite[\S2]{ct} suggests a family of hard inputs. We give each row an independent sign so that a missed row hides one bit from any estimator.

\begin{theorem}\label{thm:minimax}
Let \(1\le r\le n/4\), \(1\le\mu\le n/r\), and let each entry be observed independently with probability
\[
  0\le p\le\min\!\left\{\frac12,\frac{\mu r\log n}{16n}\right\}.
\]
For every estimator, including a randomized one, there is a fixed real \(n\times n\) matrix \(M\) of rank \(r\) and incoherence at most \(\mu\) for which
\[
  \Prob\{\widehat M\ne M\}\ \ge\ \tfrac12(1-e^{-\sqrt n/2}).
\]
The probability includes the sampling mask and the estimator's private randomness.
\end{theorem}

\begin{proof}
We first construct \(M_0=UV^{\mathsf T}\) with at least \(n/2\) nonzero rows, each supported on at most \(4n/(\mu r)\) columns. Both frames have \(r\) orthonormal columns and satisfy the required leverage bound. The following equal-leverage frame handles the case where \(n/r\) is not an integer.

Put \(a=r/n\). We construct an \(n\times r\) frame \(F\) with every squared row norm equal to \(a\). Index the columns by \(0,\ldots,r-1\), with successors taken cyclically. For \(0\le k\le r\) set
\[
 q_k=\lfloor kn/r\rfloor,\qquad t_k=k-aq_k .
\]
Thus \(0\le t_k<a\), and \(t_0=t_r=0\). For each \(k<r\), take \(c_k=q_{k+1}-q_k-2\) copies of \(\sqrt a\,e_k\) and the two rows
\[
  \sqrt{t_{k+1}/2}\,e_k \ \mathbin{\pm}\
  \sqrt{a-t_{k+1}/2}\,e_{k+1}.
\]
Here \(n/r\ge4\) ensures \(c_k\ge2\), and summing the counts gives exactly \(n\) rows. The two cross terms cancel in the column Gram matrix. Column \(k\) has squared norm
\[
 a c_k+t_{k+1}+2a-t_k
   =a(q_{k+1}-q_k)+t_{k+1}-t_k=1.
\]
For \(r=1\) the first coefficient of the pair is zero, so the same row norms and Gram identity hold. There are \(n-2r\ge n/2\) designated single-coordinate rows. Column \(k\) meets its \(c_k\) such rows and at most four rows from the adjacent pairs. Since \(ac_k\le1\), its support size is at most \(n/r+4\). The two-row orthogonalization comes from the original Spectral Tetris construction \cite[Thm.~12(ii)]{frames}.

Take \(U=F\). If \(1\le\mu<2\), also take \(V=F\). Each designated row of \(UV^{\mathsf T}\) is a nonzero multiple of a column of \(V\), with support size at most \(n/r+4\le2n/r\le4n/(\mu r)\). If \(\mu\ge2\), take the columns of \(V\) to be normalized indicators of disjoint blocks of \(s=\lceil n/(\mu r)\rceil\) entries. These fit because \(s\le n/(2r)+1\le n/r\). Their row leverages are at most \(1/s\le\mu r/n\), and \(s\le2n/(\mu r)\) since \(\mu r\le n\). This proves the claimed support and leverage bounds in both cases.

Let \(I\) be the designated rows and \(J_i\) the column support of row \(i\). Draw independent fair row signs before sampling and multiply \(U\) on the left by their diagonal matrix. Every resulting matrix has the same leverages and exactly \(r\) singular values equal to one. Row \(i\) is missed when no entry in its support is observed. This has probability \(\eta_i=(1-p)^{|J_i|}\). The inequality \(\log(1-p)\ge-2p\) for \(0\le p\le1/2\) gives \(\eta_i\ge e^{-2p|J_i|}\ge n^{-1/2}\). Distinct rows use disjoint sets of sampled entries, so
\[
 \Prob\{\text{some row of }I\text{ is missed}\}
   =1-\prod_{i\in I}(1-\eta_i)
   \ \ge\ 1-e^{-|I|/\sqrt n}
   \ \ge\ 1-e^{-\sqrt n/2}.
\]
Fix such a mask and take its least missed row in \(I\). Flipping this row's sign pairs distinct matrices with identical observations. The estimator has the same output law for each pair, which assigns total probability at most one to the two distinct answers. The average error on the pair is therefore at least \(1/2\). Averaging first over signs and masks, some fixed sign vector incurs at least the asserted error over masks and estimator randomness.
\end{proof}

When \(\mu r\log n=o(n)\), the hypotheses on \(r,\mu\) hold eventually and \(\mu r\log n/(16n)\) tends to zero. At or below this sampling probability the worst-case error stays bounded away from zero. Since the expected sample count is \(n^2p\), Theorems~\ref{thm:main} and~\ref{thm:minimax} give the sharp order \(\Theta(\mu r n\log n)\) for high-probability recovery in this square regime.

\paragraph{Relation to the lower bound of Cand\`es and Tao.}
In the rank-one example of Cand\`es and Tao \cite[\S2]{ct}, a missed row leaves one coordinate of a factor undetermined. The calculation bounds the probability that \emph{some} row is missed, and the alternative completion can depend on which row is missed. Theorem~1.7 of \cite{ct} asserts that ``there exist infinitely many pairs'' of distinct matrices whose observations agree with probability at least \(\delta\) (our \(\varepsilon\)). In \href{https://arxiv.org/pdf/0903.1476\#page=46}{\S8.1.1} the authors describe the claim as concerning ``two fixed matrices''. The following \(4\times4\) example shows that this fixed-pair conclusion can fail under the stated sampling hypothesis.

In their notation \(n\) is the side length,
\(m\) the expected number of observations, and \(p=m/n^2\) the probability of
observing each entry independently. An admissible matrix has rank at most
\(r\), with row and column leverages bounded by \(\mu_0 r/n\).
Take \(n=m=4,\ r=\mu_0=1,\ \varepsilon=2/5\). Then \(p=1/4\), and the
sampling hypothesis reduces to \((1-p)^4>2\varepsilon/n\), which holds
because \(81/256>1/5\).

Every nonzero admissible matrix has rank one. Incoherence bounds each squared coordinate of either unit singular vector by \(1/4\). The four squares sum to one, so each equals \(1/4\) and every matrix entry is nonzero. Two matrices differing at fewer than four entries agree on an entire row \(a\) and an entire column \(b\). For a nonzero matrix these determine every entry through \(M_{ij}=M_{ib}M_{aj}/M_{ab}\), while the zero matrix differs from every nonzero admissible matrix in all sixteen entries. Thus every distinct pair differs on a set \(F\) of at least four entries. A Bernoulli mask misses all of them with probability \((1-p)^{|F|}\le(3/4)^4=81/256<2/5\).

The same example extends to every \(n\ge3\). With \(m=n\), \(r=\mu_0=1\), and \(\varepsilon=2/5\), the largest collision probability of a fixed pair is \((1-1/n)^n\), attained by flipping one row of the all-ones matrix. This lies strictly between \(2\varepsilon/n\) and \(\varepsilon\), since \((1-1/n)^n\) increases from \(8/27\) to \(1/e\) for \(n\ge3\). Thus the sampling hypothesis holds and the fixed-pair conclusion fails. In Theorem~\ref{thm:minimax}, the mask determines which row sign to flip. Averaging over signs then gives each estimator one fixed input with the stated error probability.

\section{The Lean statement}\label{app:lean}

Theorem~\ref{thm:main} has been checked in Lean~4 with Mathlib. The
statement checked by the project's axiom audit is the following.

\noindent\begin{minipage}{\linewidth}
{\footnotesize
\begin{alltt}
open MatrixCompletion in
example :
    \(\exists\) C c : \(\mathbb{R}\), 0 < C \(\land\) 0 < c \(\land\)
      \(\forall\) (n\textsubscript{1} n\textsubscript{2} r : \(\mathbb{N}\)) (M : Mat n\textsubscript{1} n\textsubscript{2}) (S : CompactSVD M r) (\(\mu\) p : \(\mathbb{R}\)),
        0 < n\textsubscript{1} \(\to\) 0 < n\textsubscript{2} \(\to\) 0 < r \(\to\)
        1 \(\le\) \(\mu\) \(\to\) 0 \(\le\) p \(\to\) p \(\le\) 1 \(\to\) Incoherent S \(\mu\) \(\to\)
        C * \(\mu\) * (r : \(\mathbb{R}\)) * Real.log ((n\textsubscript{1} : \(\mathbb{R}\)) + n\textsubscript{2}) / (\(\uparrow\)(min n\textsubscript{1} n\textsubscript{2}) : \(\mathbb{R}\)) \(\le\) p \(\to\)
        1 - Real.rpow ((n\textsubscript{1} : \(\mathbb{R}\)) + n\textsubscript{2}) (-c) \(\le\) successProbability p M :=
  matrix_completion_information_limit
\end{alltt}
}
\end{minipage}
\medskip

The proof supplies \(C=2^{38}\) and \(c=3\). The formal definitions differ
in form from the text in three places.
\begin{itemize}
\item \texttt{CompactSVD M r} consists of \(U\) and \(V\) with orthonormal
columns and \(r\) positive singular values with \(M=U\Sigma V^{\mathsf T}\),
and \texttt{Incoherent} is the pair of row bounds of Section~1.
\item The nuclear norm is defined by duality, as the supremum of
\(\langle W,X\rangle\) over \(\norm W_{\op}\le1\). The formal development
proves that it is the sum of the singular values for a matrix given with a
compact singular value decomposition; Appendix~\ref{app:duality} gives the
identity in general.
\item \texttt{successProbability p M} is the sum of
\(p^{|\Omega|}(1-p)^{n_1n_2-|\Omega|}\) over the masks \(\Omega\) for
which every other matrix agreeing with \(M\) on \(\Omega\) has strictly
larger nuclear norm. It is a finite sum; no measure theory enters.
\end{itemize}

At the \href{\repo/tree/\rev}{\texttt{\rev} release}, Theorem~\ref{thm:minimax} has the Lean name
\begin{center}\small\ttfamily
MatrixCompletion.Minimax.square\_minimax\_lower\_bound
\end{center}
Given the mask and observed values, a decoder supplies a probability measure on real matrices.
Its risk sums the conditional error against the same Bernoulli mask weights as the recovery theorem.
From a fresh checkout, run \texttt{lake exe cache get} to fetch the pinned Mathlib cache.
Run \texttt{lake build} to audit both theorems against \texttt{propext},
\texttt{Classical.choice}, and \texttt{Quot.sound}.
The command \texttt{lake env lean -{}-run scripts/Replay.lean} replays both proofs through Lean's kernel.

At the same release, \texttt{bash audit/check.sh} verifies independent formulations of the \href{\auditbase/recovery/StatementBridge.lean#L133}{rectangular recovery theorem} and the \href{\auditbase/minimax/MinimaxBridge.lean#L144}{square minimax lower bound}. It also checks the \href{\auditbase/fixed-pair/CandesTaoFixedPair.lean#L158}{fixed-pair counterexample}.

\section*{Acknowledgments}

\subsection*{Human Acknowledgments}

I thank Satish Rao for his guidance on my \href{https://www2.eecs.berkeley.edu/Pubs/TechRpts/2022/EECS-2022-33.html}{Berkeley master's thesis}. That work sparked my interest in matrix completion, and our subsequent discussions encouraged me to keep working on the problem.

I thank Max Ovsiankin for checking the proof in Lean and for reading Cand\`es and Tao's proof of Theorem~1.7~\cite{ct}. He helped clarify that the alternative completion can depend on the observation mask and suggested using the missed-row construction to prove a minimax lower bound for arbitrary decoders.

I thank Sayan Paul for confirming that a Lean formalization could be developed from the manuscript alone.

\subsection*{AI Acknowledgments}

Starting at least as early as GPT-4, I asked language models for complete proofs and for help with particular calculations. The main estimates remained unresolved.

In November 2026, I asked GPT 6.2 Astra to state the recovery theorem in Lean 4. For the certificate tail starting at degree five, I assigned parallel agents to bound the signed mean, the change from revealing one entry, and the row and column variances, all under the same reveal order and stopping rule. The initial Astra runs used \texttt{high} reasoning effort, and later runs used \texttt{high} and \texttt{xhigh}.

GPT and Claude Fable 5.2 independently developed Lean arguments that applied the moment method to the full certificate and used its degree-one case to control the tangent-space sampling error. Claude also removed the reveal order from the formal argument.

Gemini 4 (Carbon and Neon) sharpened a cumulant bound and simplified the certificate analysis. GPT drafted the standalone proof from Lean, and Claude and Gemini helped clarify it. Writing out the argument led to further simplifications that were checked in Lean. Pruning unused material and simplifying the arguments reduced the Lean development from about 78,000 source lines in an early version to about 3,400 in the release, including both proofs and the axiom audit.

My principal contributions were the prior work and the direction of the proof search. I identified the off-tangent estimate as a priority, specified the bounds needed, and chose which lines of argument to pursue. The models worked out many of the arguments in the final proof, and some ideas developed over enough rounds of interaction that their origins are difficult to assign precisely.

\end{document}